\documentclass[twocolumn,showpacs,preprintnumbers,amsmath,amssymb]{revtex4}
\usepackage{graphicx}% Include figure files
\usepackage{dcolumn}% Align table columns on decimal point
\usepackage{bm}% bold math

\begin{document}

%\preprint{APS/123-QED}

\title{On the coefficients and terms of the liquid drop model \\ 
and mass formula}% Force line breaks with \\

\author{G. Royer}
\author{C. Gautier}
 \email{guy.royer@subatech.in2p3.fr}
\affiliation{Laboratoire Subatech, UMR : IN2P3/CNRS-Universit\'e-Ecole des Mines, \\
 44307 Nantes Cedex 03, France, 
}

%\affiliation{

\date{\today}% It is always \today, today,

\begin{abstract}
The coefficients of different combinations of terms of the liquid drop model have 
been determined by a least square fitting procedure to the experimental atomic 
masses. The nuclear masses can also be reproduced using  
a Coulomb radius taking into account the increase of the ratio R$_0$/A$^{1/3}$ 
 with increasing mass,
 the fitted surface energy coefficient remaining around 18~MeV.
\end{abstract}

\pacs{21.10.Dr; 21.60.Ev; 21.60.Cs}% PACS, the Physics and Astronomy
\maketitle

To predict the stability of new nuclides both in the superheavy element region and the regions close to the 
proton and neutron drip lines continuous efforts are still needed to determine the nuclear masses and therefore 
the binding energies of such exotic nuclei. Within a modelling of the nucleus by a charged liquid drop, 
semi-macroscopic models including a pairing energy have been firstly developed to reproduce the experimental nuclear masses 
\cite{wei35,bet36}. The coefficients of the Bethe-Weizs\"acker mass formula have been determined once again recently
 \cite{bas05}.   
To reproduce the non-smooth behaviour
of the masses (due to the magic number proximity, parity of the proton and neutron numbers,...) 
and other microscopic properties, 
macroscopic-microscopic approaches have been formulated, mainly the finite-range liquid drop model 
and the finite-range droplet model \cite{moll95}. Nuclear masses have also been obtained accurately within the 
statistical Thomas-Fermi model with a well-chosen effective interaction \cite{ms94rep,ms96}. Microscopic 
Hartree-Fock self-consistent calculations using mean-fields and Skyrme or Gogny forces and pairing correlations  
 \cite{samy02,sto05} as well as relativistic mean field theories \cite{bend01} have also been developed to describe 
 these nuclear masses. Finally, nuclear mass systematics using neural networks have been undertaken recently \cite{ath04}.

The nuclear binding energy B$_{nucl}$(A,Z) which is the energy necessary for separating all 
the nucleons constituting a nucleus
is connected to the nuclear mass M$_{n.m}$ by 
\begin{equation}                 
B_{nucl}(A,Z)=Zm_P+Nm_N-M_{n.m}(A,Z).
\end{equation}
This quantity may thus be easily derived from the experimental atomic masses as published in \cite{aud03} since :
\begin{equation}                 
M_{n.m}(A,Z)=M_{a.m}(A,Z)-Zm_e+B_e(Z)
\end{equation}
while the binding energy B$_e$(Z) of all removed electrons is given by \cite{lunn03}  
\begin{equation}                 
B_e(Z)=a_{el}Z^{2.39}+b_{el}Z^{5.35},
\end{equation}
with $a_{el}=1.44381\times10^{-5}$ MeV and 
$b_{el}=1.55468\times10^{-12}$ MeV.

The fission, fusion, cluster and $\alpha$ decay potential barriers are governed by 
the evolution of the nuclear binding energy with deformation. It has been shown
that four basic terms are sufficient to describe the main features of these barriers 
\cite{rr84,rr85,roy00,rm01,royzb02,rg03} :
the volume, surface, Coulomb and nuclear proximity energy terms while the introduction of 
the shell and pairing energy terms is needed to explain structure effects and 
improve quantitatively the results. Other terms have been proposed to determine accurately
 the binding energy and other nuclear characteristics : the curvature, A$^0$,
proton form factor correction, Wigner, Coulomb exchange correction,...energy terms \cite{moll95}. 

The purpose of the present work is to determine the coefficients of different combinations of terms 
of the liquid drop model by a least square fitting procedure to the experimentally available atomic 
masses \cite{aud03} and to  study whether nuclear masses can also be reproduced using, 
for the Coulomb energy, a radius which takes into account the small decrease of  the density with 
increasing mass and to determine the associated surface energy coefficient. 
The theoretical shell effects given by the Thomas-Fermi model ($7^{th}$ column of the table in \cite{ms94rep} and \cite{ms96}) 
have been selected since they 
reproduce nicely the mass decrements from fermium to $Z=112$ \cite{hof96}. They are based on the Strutinsky shell-correction
method and given for the most stable 
nuclei in the appendix. The masses of the 1522 nuclei verifying the two following conditions have 
been used : N and Z higher than 7 and the one standard deviation uncertainty on the mass lower than 100 keV 
\cite{aud03}. 

The following expansion of the nuclear binding energy has been considered
\begin{eqnarray}                 
B_{nucl}=a_v(1-k_vI^2)A-a_s(1-k_sI^2)A^{2/3}-
\frac {3}{5} \frac {e^2Z^2}{R_0} \\
+E_{pair}-E_{shell}   
-a_kA^{1/3}-a_0A^0-f_p \frac {Z^2}{A}-W \vert I \vert. \nonumber
\end{eqnarray}
The nuclear proximity energy term does not appear since its effect is effective only for necked shapes but not 
around the ground state.  
The first term is the volume energy and corresponds to the saturated exchange force and infinite nuclear matter.
In this form it includes the asymmetry energy term of the Bethe-Weizs\"acker mass formula via the relative neutron 
excess I=(N-Z)/A. The second term is the surface energy term. It takes into account the deficit of binding energy 
of the nucleons at the nuclear surface and corresponds to semi-infinite nuclear matter. The dependence on $I$ 
is not considered in the Bethe-Weizs\"acker mass formula. The third term is the Coulomb 
energy. It gives the loss of binding energy due to the repulsion between the protons. In the Bethe-Weizs\"acker mass 
formula the proportionality to Z(Z-1) is assumed.      

The pairing energy has been calculated using  
\begin{equation}
\begin{array} {ccc}
E_{pair}=-a{_p}/A^{1/2}  $ for odd Z, odd N nuclei$, \\
E_{pair}=0      $  for odd A$,  \\
E_{pair}=a{_p}/A^{1/2} $  for even Z, even N nuclei$.
\end{array} 
\end{equation}
The $a{_p}=11$ value has been adopted following first fits. Other more sophisticated expressions exist for the pairing
energy \cite{moll95,ms96}.

The sign for the shell energy term comes from the adopted definition in \cite{ms94rep}. 
It gives a contribution of $12.84~$MeV
to the binding energy for $^{208}$Pb for example.
The curvature energy a$_k$A$^{1/3}$ is a correction to the surface energy appearing
when the surface energy is considered as a function of local properties of the surface and consequently
depends on the mean local curvature.
The a$_0$A$^0$ term appears when the surface term of the liquid drop model is extended to include higher order terms in 
A$^{-1/3}$ and $I$.  
The last but one term is a proton form-factor correction to the Coulomb energy which takes into account
 the finite size of the protons. The last term is the 
Wigner energy \cite{mye77,moll95} which appears in the counting of identical pairs in a nucleus,
 furthermore it is clearly called for by the experimental masses. 

In Table 1 the improvement of the experimental data reproduction when additional terms
are added to the three basic volume, surface and Coulomb energy terms is displayed when the nuclear radius
is calculated by the formula R$_0$=r$_0$A$^{1/3}$. 

The root-mean-square deviation, defined by :
\begin{equation}
\sigma ^2= \frac {\Sigma \left \lbrack  M_{Th}- M_{Exp}\right 
\rbrack ^2}{n}
\end{equation}
has been used to compare the efficiency of these different selected sets of terms.

\begin{table*}
\caption{\label{tab:table1}Dependence of the energy coefficient values (in MeV or fm) on the selected term set
 including or not the pairing and theoretical shell 
energies and root mean square deviation. The Coulomb energy is determined by
$a_c \frac {Z^2}{A^{1/3}}$ with $a_c=3e^2/5r_0$.}
\begin{ruledtabular}
\begin{tabular}{ccccccccccccc}
$  a_v$& $k_v$ & $a_s    $ & $k_s$       &$r_0$         &  $a_k$  &$a_0$    & $f_p$    &  $W$          &$Pairing$&$Shell$&$\sigma$  \\	
\hline
15.7335  & 1.6949&  17.8048& 1.0884  &   1.2181& -          & -&   -     &         -        &n & n &2.92  \\
%\hline
15.6335 & 1.6810 &  17.2795 & 0.8840  &   1.2208 & -          & -&   -     &         -        &n& y &1.26  \\
%\hline
15.6562 & 1.6803  &  17.3492& 0.8710  &   1.2181 & -          & -&   -     &         -        &y & y &0.97  \\
%\hline
15.2374 & 1.6708 &  15.4913& 0.9223  &   1.2531& -          & 7.3742&   -     &         -        &y&y& 0.92  \\
%\hline
14.8948 & 1.6686  &  12.8138& 1.0148 &   1.2721  & 7.6659    & -&   -     &         -        & y&y&0.88  \\
%\hline
15.4443 & 1.9125 &  16.5842 & 2.4281  &   1.2444  & -          & -&   -     &     44.2135  & y&y&0.70 \\
%\hline
 15.5424& 1.7013  &  18.0625& 1.3175  &    1.2046  &     -      & - &  -1.40745&         -        & y&y& 0.68 \\
%\hline
 15.4496 & 1.8376 &  17.3525 & 2.0990 &    1.2249 &    -       & -&   -0.93794&     27.2843   &y &y& 0.584 \\
%\hline
 15.5518& 1.8452 &  18.015 & 2.0910  &   1.2176 &     -1.1207 & - &  -0.97833&         28.4963 &y&y& 0.582  \\
%\hline
 15.5293 & 1.8462 &  17.7376& 2.1163 &   1.2182 &     - & -1.5576& - 0.96851&         28.6061&y&y&0.581  \\
%\hline
 14.8497 & 1.8373 & 11.2996 & 2.7852 &   1.2505 &     20.9967 & -25.7261& - 0.6862& 26.4072&y&y&0.57  \\
\end{tabular}
\end{ruledtabular}
\end{table*}

To follow the non smooth variation of the nuclear masses with A and Z the introduction of the shell energy 
is obviously needed as well as that of the pairing term though its effect is smaller. 
The curvature term and the constant term taken alone do not allow to better fit the experimental masses
while the proton form factor correction to the Coulomb energy and the Wigner term have separately a strong effect 
in the decrease of $\sigma$. When both these two last contributions to the binding energy are taken into account
$\sigma=0.58~$MeV which is a very satisfactory value \cite{mye77,moll95,sto05}. The addition to the proton form factor 
and Wigner energy terms of the curvature energy or constant terms taken separately or together does not allow to improve 
$\sigma$. Furthermore, a progressive convergence of a$_k$ and a$_0$ is not obtained and strange surface energy coefficients
appear. Due to this strong variation and lack of stability of the curvature and constant coefficient values 
it seems preferable to neglect these terms since the accuracy is already correct without them. 
Disregarding the last line a good stability of the volume a$_v$ and asymmetry volume k$_v$ constants is observed.
The variation of the surface coefficient is larger but a$_s$ reaches a maximum of only 18~MeV. As it is 
well known the surface asymmetry coefficient k$_s$ is less easy to precise. Invariably r$_0$ has a value of around 
1.22~fm within this approach.

For the Bethe-Weizs\"acker formula the fitting procedure leads to  
\begin{eqnarray}                 
B_{nucl}(A,Z)=15.7827A-17.9042A^{2/3}\\
-0.72404 \frac {Z(Z-1)}{A^{1/3}}-23.7193I^2A 
+E_{pair}-E_{shell} \hfill \nonumber
\end{eqnarray}
with $\sigma$=1.17~MeV. That leads to r$_0$=1.193~fm and k$_v$=1.505. The non dependence of the surface energy term
on the relative neutron excess $I$ explains the $\sigma$ value.

The formula R$_0$=r$_0$A$^{1/3}$ does not reproduce the small decrease of the density with increasing
mass \cite{mye77}. In previous works \cite{rr84,rr85,roy00,rm01,royzb02,rg03} 
the formula
\begin{equation}                 
R_0=1.28A^{1/3}-0.76+0.8A^{-1/3}
\end{equation}
 proposed in Ref. \cite{blo77}
for the effective sharp radius has been retained to describe the main properties of the fusion, 
fission, cluster and $\alpha$ emission
potential barriers in the quasi-molecular shape path. It leads for example to r$_0$=1.13~fm
for $^{48}$Ca and r$_0$=1.18~fm for $^{248}$Cm. The fit of the nuclear binding energy by the expression (6) 
is reconsidered in the table 2 in calculating
the Coulomb energy by the formula $\frac {3}{5} \frac {e^2Z^2}{1.28A^{1/3}-0.76+0.8A^{-1/3}}$ without
adjustable parameter. The behaviour of the different combinations of terms is about the same as in table 1.
It seems also preferable to disregard the curvature and constant contributions. Then $\sigma$ takes the value 
0.60~MeV  to be compared to 0.58~MeV in the first table when the Coulomb energy contains an 
additional adjustable parameter.

\begin{table*}
\caption{\label{tab:table2}Dependence of the energy coefficient values (in MeV) on the selected term set including or not the pairing and shell 
energies and the corresponding root mean square deviations. The Coulomb energy coefficient is not adjusted and is determined by
$\frac {3}{5} \frac {e^2Z^2}{1.28A^{1/3}-0.76+0.8A^{-1/3}}$.}
\begin{ruledtabular}
\begin{tabular}{ccccccccccc}
 $  a_v  $&$k_v$& $a_s$     & $k_s$       & $a_k$     & $a_0$& $f_p$    & $W$           & $Pairing$ & $Shell$ & $\sigma$ \\	
\hline
15.9622  & 1.7397   &  18.0108 & 1.0627          &     -     &    -     &  - &       -        &    n      &    n    &   3.12 \\
 15.8809 & 1.7201 &  17.5366 & 0.8234 &    -      &    -     &  - &       -        &    n       &  y     &  1.56  \\
 15.8846 & 1.7256 &  17.5547 & 0.8475&   -       &  -       &  - &     -          &    y       &  y     &  1.32 \\
 15.8533 & 1.8937 &  17.2793& 1.9924 &    -      &    -& -  &    44.4714 &    y       &  y     &  1.04 \\
 15.5887 & 1.8011 &  18.194 & 1.7271 & -     &     -    & -1.98718  &      -    &    y       &  y     &  0.83  \\
 15.6089 & 1.9136 &  17.9021 & 2.4111&     -      &  - & -1.69912 &   32.1647 &    y       &  y     &  0.599 \\
 15.5833 & 1.8988 &  17.726& 2.3495 &    0.433 &   - & -1.73074&     29.8599 &    y       &  y     &  0.598  \\
 15.5996 & 1.9061&  17.8631 & 2.3757  &     -  &   0.3146 &-1.71583   &       31.0077 &    y      &  y     &  0.599 \\
 15.3737 & 1.8892 &  14.9364& 2.5745  &     12.418&    -16.7906&-1.71391&       27.8208 &    y      &  y     &  0.58 \\
\end{tabular}
\end{ruledtabular}
\end{table*}

\begin{figure}[htbp]
\begin{center}
\includegraphics[height=5.5cm]{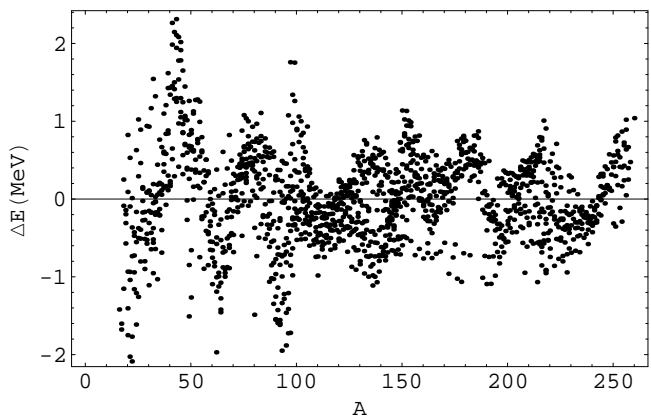}
\caption{Difference (in MeV) between the theoretical and experimental masses 
for the 1522 nuclei as a function of the mass number.}
\end{center}
\end{figure}

The figure shows that the difference between the theoretical and experimental masses never exceeds 2.25~MeV and is less
than 1.15~MeV when A is higher than 100. 
In the Generalized Liquid Drop Model \cite{rr84,rr85}
 the selected values are a$_v$=15.494~MeV, a$_s$=17.9439~MeV,  k$_v$=1.8 and k$_s$=2.6. 
They are close to the ones given $\sigma$=0.60~MeV in table 2.

As a conclusion, the most important result is that it is possible to reproduce the nuclear masses in taken a realistic formula
for the effective sharp radius given R$_0$/A$^{1/3}$=1.1~fm for the lightest nuclei and $1.18$~fm for the heaviest
ones while the surface energy coefficient remains around 18~MeV and the surface-asymmetry coefficient around 2.5.
Besides the values of the volume, surface and Coulomb energy coefficients the accuracy of the fitting depends also
 strongly on the   
selected shell and pairing energies as well as on the proton form factor and Wigner energy terms.

\appendix
\section{Table III}

\begin{table*}
\caption{\label{tab:table3}Theoretical shell energy (in MeV) extracted from \cite{ms94rep} ($7^{th}$ column) at the ground state of nuclei
 for which the half-life is higher than 1 ky.}
\begin{ruledtabular}
\begin{tabular}{cccccccccccccc}
%\hline
$^{10}B$ & $^{11}B$ &$^{12}C$  & $^{13}C$ & $^{14}C$  & $^{14}N$  &  $^{15}N$ &$^{16}O$&$^{17}O$&$^{18}O$&$^{19}F$&$^{20}Ne$& $^{21}Ne$&$^{22}Ne$\\	
 2.36& 0.74 & -0.69   &-1.03  &-0.56    &-1.33   & -0.87 &-0.45  &1.19  &1.3  & 2.76 &  2.81  &2.82  &2.19   \\
%\hline
$^{23}Na$&$^{24}Mg$&$^{25}Mg$&$^{26}Mg$&$^{26}Al$&$^{27}Al$&$^{28}Si$&$^{29}Si$&$^{30}Si$&$^{31}P$ &$^{32}S$&$^{33}S$&$^{34}S$& $^{36}S$\\	
 2.22     &  1.64 &1.77 & 0.64 & 1.89  &  0.79 &-0.26   & -0.25 &0.22  & 0.22&0.66  & 0.86  & 1.13  & 1.22   \\
%\hline
$^{35}Cl$&$^{36}Cl$&$^{37}Cl$&$^{36}Ar$&$^{38}Ar$&$^{40}Ar$&$^{39}K$&$^{40}K$&$^{41}K$&$^{40}Ca$&$^{41}Ca$&$^{42}Ca$&$^{43}Ca$&$^{44}Ca$ \\	
 1.32 & 1.48 &1.40 &1.57  &1.64  &2.45 &1.79 &2.58&2.58 &1.71  &  2.49&2.49 & 2.16 &1.7    \\
%\hline
$^{46}Ca$&$^{48}Ca$&$^{45}Sc$&$^{46}Ti$&$^{47}Ti$&$^{48}Ti$&$^{49}Ti$& $^{50}Ti$&$^{50}V$&$^{51}V$&$^{50}Cr$&$^{52}Cr$&$^{53}Cr$&$^{54}Cr$ \\	
0.69&-0.82&2.43 & 2.44 &1.95&1.44& 0.81 &-0.05  & 0.54 & -0.31& 0.74 & -0.69  &-0.38   &0.74        \\
%\hline
$^{53}Mn$&$^{55}Mn$&$^{54}Fe$&$^{56}Fe$&$^{57}Fe$& $^{58}Fe$&$^{60}Fe$&$^{59}Co$&$^{58}Ni$&$^{59}Ni$&$^{60}Ni$&$^{61}Ni$&$^{62}Ni$&$^{64}Ni$\\	
   -1.11   & 0.68      &-1.54    &0.05 & 0.72   &1.22   & 2.07   & 0.77   & -1.58   & -0.68  &-0.16    &0.59   &1.1    & 1.63     \\
%\hline
$^{63}Cu$&$^{65}Cu$&$^{64}Zn$& $^{66}Zn$&$^{67}Zn$&$^{68}Zn$&$^{70}Zn$&$^{69}Ga$&$^{71}Ga$&$^{70}Ge$&$^{72}Ge$&$^{73}Ge$&$^{74}Ge$&$^{76}Ge$\\	
1.86     &  2.33& 2.53    & 2.89    & 3.16  & 2.99    & 2.94    & 3.79    &3.71     &  4.13  & 4.08   & 4.19    &  3.82   & 2.53     \\
%\hline
$^{75}As$& $^{74}Se$&$^{76}se$&$^{77}Se$&$^{78}Se$&$^{79}Se$&$^{80}Se$&$^{82}Se$&$^{79}Br$&$^{81}Br$&$^{80}Kr$&$^{81}Kr$&$^{82}Kr$& $^{83}Kr$\\	
 4.08    & 4.42     &  4.08   & 4.06    & 3.27    & 2.87    &  1.89   &   0.38  &   4.07  &   2.28  &   4.39  &  3.77 & 2.74    & 1.65       \\
%\hline
$^{84}Kr$&$^{86}Kr$&$^{85}Rb$&$^{87}Rb$&$^{86}Sr$&$^{87}Sr$&$^{88}Sr$&$^{89}Y$&$^{90}Zr$&$^{91}Zr$&$^{92}Zr$& $^{93}Zr$&$^{94}Zr$&$^{96}Zr$\\	
0.96   & -0.40   & 1.13    & -0.35   &  0.79   &  0.05   & -0.97   &  -1.19 &  -1.63  & -0.47 & 0.46    &   1.53   &  2.54   &   3.49   \\
%\hline
$^{92}Nb$&$^{93}Nb$&$^{94}Nb$&$^{92}Mo$&$^{93}Mo$&$^{94}Mo$&$^{95}Mo$&$^{96}Mo$&$^{97}Mo$& $^{98}Mo$&$^{100}Mo$&$^{97}Tc$&$^{98}Tc$&$^{99}Tc$\\	
 -0.57  &   0.44  &  1.51   &  -2.12  &   -1.07 & -0.12   & 0.97    &  1.79 &  2.46    &    2.98 &  3.62    & 1.26    & 2.05    & 2.64    \\
%\hline
$^{96}Ru$&$^{98}Ru$&$^{99}Ru$&$^{100}Ru$&$^{101}Ru$&$^{102}Ru$&$^{104}Ru$& $^{103}Rh$&$^{102}Pd$&$^{104}Pd$&$^{105}Pd$&$^{106}Pd$&$^{107}Pd$&$^{108}Pd$\\	
-1.11     & 0.57    &   1.36  &  2.00    &   2.54 &   2.98&3.49     & 2.44     &  0.63    & 1.83     & 2.39    & 2.80  &  3.08    & 3.34       \\
%\hline
$^{110}Pd$&$^{107}Ag$&$^{109}Ag$&$^{106}Cd$&$^{108}Cd$& $^{110}Cd$&$^{111}Cd$&$^{112}Cd$&$^{113}Cd$&$^{114}Cd$&$^{116}Cd$&$^{113}In$&$^{115}In$&$^{112}Sn$\\	
  3.42   &  2.20   &  2.91     &  0.31 &1.35     & 2.11      &  2.42    & 2.52     & 2.61     & 2.50     &   2.26   &  1.83    & 1.97     &  0.37     \\
%\hline
$^{114}Sn$&$^{115}Sn$&$^{116}Sn$& $^{117}Sn$&$^{118}Sn$&$^{119}Sn$&$^{120}Sn$&$^{122}Sn$&$^{124}Sn$&$^{126}Sn$&$^{121}Sb$&$^{123}Sb$&$^{120}Te$&$^{122}Te$\\	
0.81    &  1.03  & 0.94     & 0.96      &  0.75    & 0.70     & 0.19     & -0.99    & -2.51    &  -4.36   &  0.76    & -0.16    & 2.08     & 1.55   \\
%\hline
$^{123}Te$& $^{124}Te$&$^{125}Te$&$^{126}Te$&$^{128}Te$&$^{130}Te$&$^{127}I$&$^{129}I$&$^{128}Xe$&$^{129}Xe$&$^{130}Xe$&$^{131}Xe$&$^{132}Xe$& $^{134}Xe$\\	
 1.30     & 0.66      &  0.25    & -0.62    & -2.46    & -4.74    &0.35    & -1.29   &   1.01   & 0.48    & -0.31    &  -1.11 & -2.36   & -4.92   \\
%\hline
$^{136}Xe$&$^{133}Cs$&$^{135}Cs$&$^{132}Ba$&$^{134}Ba$&$^{135}Ba$&$^{136}Ba$&$^{137}Ba$&$^{138}Ba$&$^{137}La$&$^{138}La$& $^{139}La$&$^{136}Ce$&$^{138}Ce$\\	
  -7.2   &   -1.28  &  -3.90   &  0.93    &  -0.55   &   -1.45  &   -3.01  & -4.18   &  -5.29     &  -2.22&   -3.37  &   -4.50   &  0.70    & -1.57      \\
%\hline
$^{140}Ce$&$^{142}Ce$&$^{141}Pr$&$^{142}Nd$&$^{143}Nd$&$^{144}Nd$&$^{145}Nd$&$^{146}Nd$&$^{148}Nd$& $^{150}Nd$&$^{144}Sm$&$^{146}Sm$&$^{147}Sm$&$^{148}Sm$\\	
 -3.86    & -2.06    & -3.26    & -2.88    &  -2.14   & -1.04    & 0.10    &  0.56 & 0.85   &  0.54     &  -2.28   & -0.41    & 0.67     &  1.12    \\
%\hline
$^{149}Sm$&$^{150}Sm$&$^{152}Sm$&$^{154}Sm$&$^{151}Eu$&$^{153}Eu$&$^{150}Gd$& $^{152}Gd$&$^{154}Gd$&$^{155}Gd$&$^{156}Gd$&$^{157}Gd$&$^{158}Gd$&$^{160}Gd$\\	
1.22    &  1.31    &   0.90    & 0.38     &  1.39     &  1.02 &   1.30    & 1.59      &  1.33    & 1.05     & 0.89     & 0.62     &  0.56    &    0.21  \\
%\hline
$^{159}Tb$&$^{154}Dy$&$^{156}Dy$&$^{158}Dy$&$^{160}Dy$& $^{161}Dy$&$^{162}Dy$&$^{163}Dy$&$^{164}Dy$&$^{165}Dy$&$^{163}Ho$&$^{165}Ho$&$^{162}Er$&$^{164}Er$\\	
 0.64   & 1.63     &   1.56   & 1.24 & 0.92     &   0.64    &    0.47  &    0.16  &   -0.06  & -0.41    &  0.46     &  -0.12   &  1.20    &   0.70      \\
%\hline
$^{166}Er$&$^{167}Er$&$^{168}Er$& $^{170}Er$&$^{169}Tm$&$^{168}Yb$&$^{170}Yb$&$^{171}Yb$&$^{172}Yb$&$^{173}Yb$&$^{174}Yb$&$^{176}Yb$&$^{175}Lu$&$^{176}Lu$\\	
 0.07    &  -0.37 &  -0.54    &   -1.08    &   -0.60  &   0.32   &   -0.34   & -0.76    &  -0.94   &   -1.32  &  -1.30   &  -1.74   &  -1.23    &   -1.62  \\
%\hline
$^{174}Hf$& $^{176}Hf$&$^{177}Hf$&$^{178}Hf$&$^{179}Hf$&$^{180}Hf$&$^{182}Hf$&$^{181}Ta$&$^{180}W$&$^{182}W$&$^{183}W$&$^{184}W$&$^{186}W$& $^{185}Re$\\	
 -0.38    &   -0.90   &    -1.33 &  -1.53   & -1.97    &  -1.99   & -2.16    &   -2.02  &   -1.21 &   -1.71 &  -2.00  &  -2.02 &  -2.38   &   -2.19    \\
%\hline
$^{187}Re$&$^{184}Os$&$^{186}Os$&$^{187}Os$&$^{188}Os$&$^{189}Os$&$^{190}Os$&$^{192}Os$&$^{191}Ir$&$^{193}Ir$&$^{190}Pt$& $^{192}Pt$&$^{194}Pt$&$^{195}Pt$\\	
 -2.48 &  -1.61   &   -1.88  & -2.16     &  -2.08   &  -2.43   &   -2.47  &  -3.50   &  -2.54   &  -3.62 &  -0.97  &   -2.01   &  -3.31    &   -4.04   \\
%\hline
$^{196}Pt$&$^{198}Pt$&$^{197}Au$&$^{196}Hg$&$^{198}Hg$&$^{199}Hg$&$^{200}Hg$&$^{201}Hg$&$^{202}Hg$& $^{204}Hg$&$^{203}Tl$&$^{205}Tl$&$^{202}Pb$&$^{204}Pb$\\	
   -4.80  &   -6.11  &   -5.56  & -4.51    &   -5.99  &   -6.75  &    -7.52  &   -8.37& -9.11    &   -10.69  &  -9.97   &    -11.58 &   -8.22  & -10.02  \\
%\hline
$^{205}Pb$&$^{206}Pb$&$^{207}Pb$&$^{208}Pb$&$^{208}Bi$&$^{209}Bi$&$^{226}Ra$& $^{229}Th$&$^{230}Th$&$^{232}Th$&$^{231}Pa$&$^{233}U$&$^{234}U$&$^{235}U$\\	
   -11.00 &   -11.82 &   -12.68 &  -12.84  &  -11.70  &  -11.95 &  -0.30    &   -0.52   &  -0.43   & -0.60    &   -0.79  &  -1.27  &   -1.23 &  -1.46 \\
%\hline
$^{236}U$&$^{238}U$&$^{236}Np$&$^{237}Np$&$^{239}Pu$& $^{240}Pu$&$^{242}Pu$&$^{244}Pu$&$^{243}Am$&$^{245}Cm$&$^{246}Cm$&$^{247}Cm$&$^{248}Cm$&$^{247}Bk$\\	
-1.30  &   -1.27  &   -1.85  &  -1.74 & -2.12    &    -1.95  &   -1.99  &   -2.08  &   -2.44  &  -3.05    &   -2.96  &   -3.17   &    -3.00 & -3.46      \\
%\hline
\end{tabular}
\end{ruledtabular}
\end{table*}

\end{document}